\documentclass[%
  reprint,
  twocolumn,
 showpacs,
 showkeys,
 preprintnumbers,
 amsmath,amssymb,
 aps,
 pra,
  longbibliography,
 ]{revtex4-1}

\usepackage[dvipsnames]{xcolor}

\usepackage{mathptmx}% http://ctan.org/pkg/mathptmx

\usepackage{amssymb,amsthm,amsmath,bm}

\usepackage{tikz}
\usetikzlibrary{calc,decorations.pathreplacing,decorations.markings,positioning,shapes}

\usepackage[breaklinks=true,colorlinks=true,anchorcolor=blue,citecolor=blue,filecolor=blue,menucolor=blue,pagecolor=blue,urlcolor=blue,linkcolor=blue]{hyperref}
\usepackage{graphicx}% Include figure files
\usepackage{url}

\begin{document}

\title{Quantum violation of the Suppes-Zanotti inequalities and ``contextuality''}

\author{Karl Svozil}
\email{svozil@tuwien.ac.at}
\homepage{http://tph.tuwien.ac.at/~svozil}

\affiliation{Institute for Theoretical Physics,
TU Wien,
\\
Wiedner Hauptstrasse 8-10/136,
1040 Vienna,  Austria}

\date{\today}

\begin{abstract}
The Suppes-Zanotti inequalities involving the joint expectations of just three binary quantum observables are (re-)derived by the hull computation of the respective correlation polytope.
A min-max calculation reveals its maximal quantum violations correspond to a generalized Tsirelson bound.
Notions of ``contextuality'' motivated by such violations are critically reviewed.
\end{abstract}


\maketitle

\section{Two-partite vector-based expectations not satisfying classical bounds}

Classical bounds on probabilities and expectations can be expected to be ``violated'' by or ``being different'' from quantum probabilities and expectations because the latter are based on multi-dimensional vectorial entities
whereas the former are based on scalars in (sub)sets of power sets.
Exactly how these violations are operationalized and measured has developed from an intuitive, heuristic search in the early days~\cite{bell,chsh,wigner-70}
into a systematic method~\cite{froissart-81,Garg1984,pitowsky-86,cirelson,svozil-2017-b}.

The most elementary expression of the classical versus quantum difference quoted earlier
is the two-partite correlation function of two dichotomic observables $X,Y\in \{-1,+1\}$.
It is empirically collected from a series of $N$ measurements of $X$ and $Y$ and defined by
$\langle X,Y \rangle_s  \approx  \frac{1}{N}\sum_{i=1}^N   X_i Y_i$, where the index $i$ refers to the $i$th measurement,
and $s$ refers to a specific (unaltered) state on which these repetitive measurements are performed.
It is assumed that, if $N$ increases, the limit exists and is monotonically approached --
that is, for ``large enough'' $N$, $\langle X,Y \rangle_s$ is a ``good approximation''~\cite{Uffink2011-UFFSPS}.

\subsection{Classical predictions on ``singlet-type'' states}

It is not too difficult to model a classical two-partite state $q$ which shows ``singlet-like'' characteristics
-- an example would be the angular momentum in a particular spacial direction
of two fragments of a bomb which originally had no angular momentum in any direction~\cite{peres}.
An argument involving equidistribution of angular momenta of the fragments reveals a linear classical correlation on such a state; that is,
$\langle X,Y \rangle_c  = \textsf{\textbf{E}}(X,Y) = \textsf{\textbf{E}}(\theta) = -1+2 \theta    /\pi$,
where the angle $0 \le \theta \le \pi$ characterizes the ``spatial separation'' of the directions of these observables $X$ and $Y$.

\subsection{General classical predictions}

To construct a generic classical situation, a generalized urn model~\cite{wright} is introduced
which can also be phrased in terms of finite-state identification problems of automata allowing
complementarity~\cite{e-f-moore}.
Its formalization is in terms of set-theoretic partitions~\cite{svozil-2001-eua} and power sets.

In terms of generalized urn models, we consider urns filled with black balls painted with three different colors,
one color per observable $X$, $Y$, and $Z$.
Since each observable may have two different outcomes we can, for instance, label these outcomes by ``$+$'' and ``$-$'',
printed on these balls in the respective colors.
There are eight such ball types. As the urn is filled with an arbitrary distribution of ball types, it can only be ascertained that
they occur with probabilities $0\le \lambda_{\pm \pm \pm} \le 1$, were the indices refer to the respective
symbols in the colors associated with our three observables.
Since in such a scheme the ball types are mutually exclusive and their enumeration is complete (i.e., exhaustive),
we can suppose that $\sum_{i,j,k \in \{+,-\}}\lambda_{ijk}=1$, and
the joint expectations add up accordingly; e.g.,
$\textsf{\textbf{E}}(X,Y)
= \sum_{k \in \{+,-\}}\left[ \lambda_{++,k} + \lambda_{--,k} - \left( \lambda_{+-,k} + \lambda_{-+,k}\right)\right]$.

\subsection{Quantum predictions on a singlet state}

The quantum predictions of a single observable in an arbitrary direction
characterized by the spherical coordinates $0\le \theta \le \pi$ and $0\le \varphi < 2\pi$
is derived from the Pauli spin matrices $\sigma_x$,  $\sigma_y$ and $\sigma_z$
forming the spin operator
${\bm{\sigma}} (\theta , \varphi ) = \sigma_x \sin\theta \cos\varphi  + \sigma_y \sin\theta \sin\varphi  + \sigma_z \cos\theta$
and the single particle projection operator
$\textsf{\textbf{S}}_\pm (\theta , \varphi) =\frac{1}{2} \left[ \mathbb{I}_2  \pm \bm{\sigma} (\theta , \varphi ) \right]$
for the states ``$-$'' and ``$+$'', respectively.
The respective two-partite projection operators are
$\textsf{\textbf{S}}_{\pm_1\pm_2} (\theta_1 , \varphi_1, \theta_2 , \varphi_2) = \textsf{\textbf{S}}_{\pm_1} (\theta_1 , \varphi_1) \otimes  \textsf{\textbf{S}}_{\pm_2} (\theta_2 , \varphi_2)$.
Finally, the operator associated with the two-partite expectations is
$\textsf{\textbf{F}}(X,Y) = \textsf{\textbf{F}}(\theta_1 , \varphi_1, \theta_2 , \varphi_2) =
\textsf{\textbf{S}}_{++}+
\textsf{\textbf{S}}_{--}-
\left(
\textsf{\textbf{S}}_{+-}+
\textsf{\textbf{S}}_{-+}
\right)={\bm{\sigma}} (\theta_1 , \varphi_1 ) \otimes {\bm{\sigma}} (\theta_2 , \varphi_2 )$.

Suppose we are interested in the correlation function for a singlet state in the Bell basis
$q = \vert \Psi_- \rangle \langle \Psi_-\vert $
with
$\vert \Psi_- \rangle = \frac{1}{\sqrt{2}}\begin{pmatrix}0,1,-1,0\end{pmatrix}^\intercal$,
then the quantum prediction yields
$\langle X,Y \rangle_q  = \textsf{\textbf{F}}(\theta_1,\varphi_1,\theta_2,\varphi_2) =
 - \left[\cos\theta_1 \cos\theta_2 + \cos(\varphi_1 - \varphi_2) \sin\theta_1 \sin\theta_2 \right]$.
For $\varphi_1=\varphi_2$ this reduces to the well-known cosine form
$\langle X,Y \rangle_q  =\textsf{\textbf{F}}(\theta_1,0,\theta_2,0) =
\textsf{\textbf{F}}(\theta_1-\theta_2) = - \cos (\theta_1 - \theta_2)$,
that is, the two-partite correlation for dichotomic observables $X,Y=\pm 1$ of the two-partite singlet state
is proportional to the Euclidean scalar product between the vectors associated with $X$ and $Y$.

The maximal quantum-to-classical violations
\begin{equation}
\max_{\theta \in \{0, \pi\}} \left|\textsf{\textbf{E}}(\theta ) -\textsf{\textbf{F}}(\theta )\right|
=
\sqrt{1-\left(\frac{2}{\pi}\right)^2}-\frac{2}{\pi }\cos^{-1}\frac{2}{\pi }\approx 0.2
\end{equation}
resulting from less,
as well as more,
equal occurrences of the joint observables $++$/$--$ and $+-$/$-+$,
occur at angles $(d/d\theta )\left[ \textsf{\textbf{E}}(\theta ) -\textsf{\textbf{F}}(\theta ) \right] = 0$,
that is, at
\begin{equation}
\begin{aligned}
\theta &=\sin ^{-1} \frac{2}{\pi}\text{ as well as}\\
\theta &= \pi - \sin ^{-1} \frac{2}{\pi} \text{, respectively.}
\end{aligned}
\end{equation}

% N[ArcSin[2/Pi]]; Plot[{-Cos[t], -1 + 2 t/Pi}, {t, 0, Pi}]; D [Cos[t] - 1 + 2 t/Pi, t]; Plot[2/Pi - Sin[t], {t, 0, Pi}]; TeXForm  [TrigExpand[-1 + (2 ArcSin [2/Pi])/Pi + Cos[ArcSin[2/Pi]]]]

\subsection{Quantum predictions on more general pure states}

By a min-max calculation~\cite{filipp-svo-04-qpoly-prl} it is not too difficult to compute those quantum states which,
given arbitrary angles between the two observables $X$ and $Y$, yield the minimal and maximal correlations:
all that is needed is the eigensystem  of $\textsf{\textbf{F}}(\theta_1 , \varphi_1, \theta_2 , \varphi_2)$.
Rather than enumerating this eigensystem in full generality the special case $\theta_1=\theta$ and $\theta_2 = \varphi_1 = \varphi_2 =0$ is posted,
resulting in the (decomposable) vectors (modulo normalization)
\begin{equation}
\begin{aligned}
\vert \psi_\text{1,min}\rangle &= \begin{pmatrix}
0,  {\cos \theta +1},0,{\sin \theta}
\end{pmatrix}^\intercal \text{ as well as}\\
\vert \psi_\text{2,min}\rangle &= \begin{pmatrix}
 {\cos \theta -1},0 ,{\sin \theta},0
\end{pmatrix}^\intercal
\end{aligned}
\end{equation}
for the minimal expectation $\langle X,Y \rangle  = -1$;
and
\begin{equation}
\begin{aligned}
\vert \psi_\text{1,max}\rangle &= \begin{pmatrix}
0,  {\cos \theta -1},0,{\sin \theta}
\end{pmatrix}^\intercal\text{ as well as}\\
\vert \psi_\text{2,max}\rangle &= \begin{pmatrix}
 {\cos \theta +1},0 ,{\sin \theta},0
\end{pmatrix}^\intercal
\end{aligned}
\end{equation}
for the maximal expectation $\langle X,Y \rangle  = 1$,
respectively.

\section{The case of tree observables}
\subsection{Classical bounds}

One might as well stop here, contemplate the elementary difference between two forms of probabilities based on scalars and power sets in the classical case,
and on vectors and the vector space spanned by them in the quantum case,
and leave it at that.
However, this is not what happened historically: Bell and others tried to find criteria for non-compliance with classical behavior involving more than just two observables.
In particular, Suppes and Zanotti~\cite{Suppes-81,Brody-1989,Khrennikov2020} presented  special cases of what Boole called ``conditions of possible experience''~\cite{Boole-62,Pit-94}
involving just three dichotomic observables $X,Y,Z\in \{-1,+1\}$.

The original method of deriving these bounds is rather involved.
But with today's convex polytope techniques~\cite{froissart-81,pitowsky,cirelson} it is not too difficult to derive those inequalities:
(i) form all possible combinations of joint occurrences by multiplying the respective dichotomic observables --
in this case
$\textsf{\textbf{E}}(X,Y) = XY$,
$\textsf{\textbf{E}}(X,Z) = XZ$,
$\textsf{\textbf{E}}(Y,Z) = YZ$;
(ii) form the 3-tuples (that is, the finite ordered list or sequence) of all three numbers for particular instances of $X,Y,Z \in \{-1,+1\}$
$ \begin{pmatrix}
\textsf{\textbf{E}}(X,Y) , \textsf{\textbf{E}} (X,Z) , \textsf{\textbf{E}} (Y,Z)
\end{pmatrix}
=
\begin{pmatrix} XY , XZ , YZ \end{pmatrix} $,
(iii) pretend these 3-tuples are coordinates (with respect to the Cartesian three-dimensional standard basis)
of vertices of a convex polytope,
and
(iv) according to the Minkowski-Weyl ``main'' representation theorem~\cite{ziegler,Schrijver,Fukuda-techrep}
represent this polytope as its facets by the hull computation~\cite{Fukuda-techrep,Pit-91}.
These facet (in)equalities represent Boole-Bell type ``conditions of possible (classical) experience''.

With three dichotomic observables, such procedures result in eight three-dimensional row vectors.
Four of them are linearly independent.
They are interpreted as the vertices of a correlation polytope.
The row vectors, stacked on top of one another, form a
 $4 \times 3$ Travis~\cite{travis-mt-62} matrix~\cite{greechie-66-PhD}
\begin{equation}
T_{ij}=\begin{pmatrix}
   +1  & +1 &  +1  \\
   +1  & -1 &  -1 \\
   -1  & +1 &  -1 \\
   -1  & -1 &  +1
\end{pmatrix} .
\end{equation}
The hull computation
(eg, by  {\tt pycddlib}~\cite{pycddlib}, a Python wrapper of Fukuda's {\tt cddlib} algorithm~\cite{cdd-pck}
implementing  the Double Description Method~\cite{MRTT53})
yields the four Suppes-Zanotti-Brodi inequalities~\cite{Suppes-81,Brody-1989}
\begin{equation}
\begin{split}
- 1  \le   \textsf{\textbf{E}}(X,Y) + \textsf{\textbf{E}}(X,Z) + \textsf{\textbf{E}}(Y,Z)   ,   \\
- 1  \le   - \textsf{\textbf{E}}(X,Y) - \textsf{\textbf{E}}(X,Z) + \textsf{\textbf{E}}(Y,Z) ,   \\
- 1  \le   \textsf{\textbf{E}}(X,Y) - \textsf{\textbf{E}}(X,Z) - \textsf{\textbf{E}}(Y,Z)   ,   \\
- 1  \le   -\textsf{\textbf{E}}(X,Y) + \textsf{\textbf{E}}(X,Z) - \textsf{\textbf{E}}(Y,Z)
.
\end{split}
\label{2020-ex-SZI}
\end{equation}

\subsection{Quantum bounds by min-max calculation}

The min-max calculations~\cite{filipp-svo-04-qpoly-prl}
of the associated operators
$
\textsf{\textbf{F}}(X,Y) \pm \textsf{\textbf{F}}(X,Z) \pm \textsf{\textbf{F}}(Y,Z)
$
with the quantum expectation $\textsf{\textbf{F}}$ as defined earlier
amounts to summing up the separate terms and determining the eigensystem of these new observables.
It yields quantum bounds allowing ranges bounded by
\begin{equation}
-3 <  \textsf{\textbf{F}}(X,Y) \pm \textsf{\textbf{F}}(X,Z) \pm \textsf{\textbf{F}}(Y,Z)   < 3
\label{2020-ex-SZIq}
\end{equation}
which violate the classical ones~(\ref{2020-ex-SZI})
by almost the greatest algebraically possible amount.

For the sake of more concrete realizations, we shall set all azimuthal angles to zero and take equidistant polar angles
such that the directions of $X$, $Y$, and $Z$ in configuration space are $0$, $\theta$, and $2 \theta$, respectively.
Then the min-max computation associated with
$\textsf{\textbf{F}}(0,\theta ) + \textsf{\textbf{F}}(0,2 \theta ) + \textsf{\textbf{F}}(\theta , 2 \theta )$
exhibits two eigenvalues
\begin{equation}
\mu_1= -(5 + 4 \cos\theta)^{1/2} \le  -(1 + 2 \cos \theta )=\mu_2
\end{equation}
which, in a certain domain of $\theta$, violate the first inequality in~(\ref{2020-ex-SZI}).
The associated pure states are proportional to
% TeXForm[FullSimplify[   TrigExpand[    FullSimplify[      TrigReduce[       Eigensystem[SZ1[0, 0, \[Theta], 0, 2 \[Theta], 0]][[2,         3]]]]]*2 (1 + Cos[\[Theta]])]]
\begin{equation}
\begin{aligned}
& \vert {\bf x}_1 \rangle =
\begin{pmatrix}
a ,
b  ,
 -b  ,
a
\end{pmatrix}^\intercal
\text{, where}\\
&\quad a  = 2 (\cos \theta +1)\sin \theta \text{ and } \\
&\quad b =  2 \cos \theta +\cos (2 \theta )+\sqrt{5+4 \cos \theta} \text{, as well as }\\
% TeXForm[FullSimplify[   TrigExpand[    FullSimplify[      TrigReduce[       Eigensystem[SZ1[0, 0, \[Theta], 0, 2 \[Theta], 0]][[2,         1]]]]]*2 (1 + Cos[\[Theta]])*Sin[\[Theta]]]]
&\vert {\bf x}_2 \rangle =        \begin{pmatrix}
-\sin  \theta  ,\cos  \theta  ,\cos  \theta  ,\sin  \theta
\end{pmatrix}^\intercal
\text{, respectively.}
\end{aligned}
\end{equation}
Note that for $\theta \rightarrow 0$ these two states converge to indecomposable vectors proportional to the  Bell basis states
$
\begin{pmatrix}
0 , 1  , -1 , 0
\end{pmatrix}^\intercal
$
as well as
$
\begin{pmatrix}
0  ,1  ,1  ,0
\end{pmatrix}^\intercal
$.
Indeed,   for  $\theta \rightarrow 0$,
the two other eigenstates rendering the two eigenvalues
$(5 + 4 \cos\theta )^{1/2} ,
1 + 2 \cos \theta  \rightarrow 3$, converge to the remaining states in the Bell basis.

\subsection{Composition of higher-order distribution by lower-order ones}

For some ``practical'' application recall
Specker's story about~\cite{specker-60} {\em ``a wise man from Ninive $\ldots$ who was $\ldots$ concerned almost exclusively about his daughter''}
and an oracle potential suitors had to cope with:
{\em ``The suitors were led in front of a table on which three boxes were positioned in a row, and they were ordered to indicate which of the boxes contained a gem and which were empty.
And now no matter how many times they tried, it seemed to be impossible to solve the task.
After their predictions, each of the suitors was ordered to open two boxes which they had indicated to be both empty or both not empty:
it turned out each time that one contained a gem and the other did not, and,
to be precise, sometimes the gem was in the first, sometimes in the second of the boxes that were opened.
But how can it be possible that from three boxes neither two can be indicated as empty, nor as not empty?''}

A similar scheme was mentioned by Garg and Mermin~\cite{Garg1984}:
{\em ``if we have three dichotomic
variables each of which assumes either the value 1 or -1 with equal
probability and all the pair distributions vanish unless the members of the
pair have different values~$\ldots$~.''}

These scenarios mention three observables and strict anti-correlations between pairs of observable outcomes, such that
$ \textsf{\textbf{E}}(X,Y) =\textsf{\textbf{E}}(X,Z) = \textsf{\textbf{E}}(Y,Z)=-1 $.
As can be readily checked by the (maximal) violation of the first Suppes-Zanotti-Brodi inequalities~(\ref{2020-ex-SZI}) no classical global probability distribution allows this.
But quantum mechanics can ``almost'' provide a realization as it yields ``almost perfect'' anti-correlations
at ``almost vanishing'' angles $0< \theta \ll 1$.
The ``reason'' for this is threefold: (i) the quantum expectation function, as mentioned earlier, is
$\langle X,Y \rangle_q  =\textsf{\textbf{F}}(\theta_1,0,\theta_2,0) = - \cos (\theta_1 - \theta_2)$;
(ii) the three expectation functions are complementary and therefore cannot be measured simultaneously -- they have no simultaneous value definiteness; and
(iii) the quantum resources exploit a four-dimensional Hilbert space with probabilities based on vectors rather than scalars.

It might be worth noting that Greenberger, Horne, and Zeilinger proposed another, adaptive, protocol involving expectations of order three and
going beyond stochastic quantum violations of classical predictions~\cite{ghz,ghsz,mermin}
which could be rewritten as a game ``people play''~\cite{PhysRevLett.82.1345,panbdwz,bacon-ghzgames-2006}
in which particular quantum states allow certain players always to win whereas this is not guaranteed classically~\cite{svozil-2020-ghz}.

\section{The case of four and more observables}

For completeness, we just mention that the addition of an additional variable yields the well-known Clauser-Horne-Shimony-Holt inequalities~\cite{chsh}.
A polytope derivation can be found in Refs.~\cite{froissart-81,pitowsky,cirelson}.
Its quantum bound $-2\sqrt{2} \le \textsf{\textbf{F}}(W,Y) + \textsf{\textbf{F}}(W,Z) + \textsf{\textbf{F}}(X,Y)  - \textsf{\textbf{F}}(X,Z) \le
2\sqrt{2}$ derived by Cirel'son (aka Tsirelson)~\cite{cirelson:80} can be straightforwardly obtained from a min-max calculation~\cite{filipp-svo-04-qpoly-prl}
of its eigensystem. The quantum states rendering this bound can be represented by the vectors proportional to
$\begin{pmatrix} -1, 1, 1, 1\end{pmatrix}^\intercal$
and $\begin{pmatrix} -1, -1, -1, 1\end{pmatrix}^\intercal$, respectively.

The polytope method can be straightforwardly scaled to derive Boolean ``bounds of classical experience''
for over four observables~\cite{2000-poly,sliwa-2003,collins-gisin-2003}.
Their respective quantum violations can again be derived by a min-max calculation~\cite{filipp-svo-04-qpoly-prl}.

\section{``Contextuality'' in context}

Let me add a cautionary remark on the widely held opinion that violations of classical Boolean criteria
such as the Suppes-Zanotti-Brodi inequalities
suggest or even imply ``contextuality''.
Presently the term ``contextual'' is often heuristically used as ``violation of some inequality
that is derived by assuming classical probability distributions''~\cite{cabello:210401,cabello2020converting}.
There are a variety of notions~\cite{svozil:040102}
and accompanying measures~\cite{svozil-2011-enough,Abramsky-2017,Khrennikov2017,KujalaDzhafarov-2019} for the term ``contextuality''.

This ``modern'' quantitative use of the word can be contrasted with Bohr's synthetic suggestion of a {\em conditionality of phenomena} by~\citet{bohr-1949}
{\it ``the impossibility of any sharp separation between the behavior of atomic
objects and the interaction with the measuring instruments which serve to define the conditions
under which the phenomena appear.''}
A related proposition from the realist Bell contends that~\cite{bell-66}
{\it ``the result of an observation may reasonably depend $\ldots$ on the complete disposition of the apparatus.''}

However, this does not imply -- and it may be even misleading to believe --
that these conceivable ``results of an observation'' (aka outcome/event) are ``dormant'' properties of the object (alone) which become ``visible/actuated''
by some ``complete disposition of the apparatus'' (aka context).
More precisely, there need not be any functional
(in the sense of uniqueness) dependency of the outcome which originates in causes or factors within the observed system;
no value definite intrinsic property of the object alone.
One could understand Bohr and Bell also by their insistence that the value definite properties (characterizing its physical state) of the object
become ``amalgamated'' with (properties of) the measurement apparatus,
so that an observation signals the combined information both of the object as well as of the measurement apparatus.

If one prepares a quantized system to be in a pure state formalized by a vector,
then it is perfectly value definite for observable properties corresponding to that same preparation (context).
But if there is a mismatch between preparation and measurement, the latter environment distorts value definiteness by an ``inflow''
of information from ``outside of'' the object.
Consequently, it makes no sense to speak of any such measurement result as ``being an element of physical reality'' associated with the observed system alone --
one has to add the (open) environment which ``translates'' the preparation into the measurement, thereby introducing (external with respect to the object) noise~\cite{svozil-2003-garda}.

\section{What propositions support which probabilities?}

For comparing probabilities and expectations on propositional structures
I maintain that in all such considerations two issues need to be distinguished as separate criteria:
\begin{itemize}
\item[(i)]
Given some particular type of propositional structure (aka logics); which variety of probability distribution(s) is(are) supported by
this propositional structure?
\item[(ii)]
Given two or more such varieties of probability distributions, exactly what types of probability distributions should be compared with one another?
Is this not a question that needs to be settled for the particular type of systems dealt with?
\end{itemize}

I am unaware of any systematic way of answering the first question (i).
One approach, motivated by
Gleason-type theorems~\cite{Busch-2003,caves-fuchs-2004,Granstrom-mt,wright-Victoria,Wright_2019,Wright2019},
is in terms of  is Cauchy-type functional equations.

For instance, the same propositional structure may,
on the one hand, support a classical hidden variable theory
based on scalars as well as on (subsets of) a single Boolean algebra,
while
on the other hand, accommodate a quantum interpretation based on multi-dimensional vector space entities~\cite{svozil-2018-b}.
Take, for example, the Specker bug/cat's cradle~\cite{kochen2,Pitowsky2003395,pitowsky-06},
or the house/pentagon/pentagram~\cite{greechie-1974,wright:pent,Klyachko-2008} logics:
both have a classical interpretation in terms of partitions of the sets of two-valued measures~\cite{svozil-2001-eua}
as well as a faithful orthogonal representation~\cite{lovasz-79,lovasz-89} as vectors.

But there are also structures that do not allow any global classical probability distribution
yet support a vector coordinatization (aka faithful orthogonal representation).
Examples are the Specker bug combo denoted by $\Gamma_3$ by Kochen and Specker~\cite{kochen1}
that has a nonseparable set of two-valued states.
In the extreme case there exists no classical truth assignment (relative to admissibility; ie, exclusivity and completeness):
take, for example, $\Gamma_2$~\cite{kochen1}, or the logics introduced in Refs.~\cite{cabello-96,2015-AnalyticKS}.
One ``demarcation criterion'' is the separability of the observables by two-valued states, as expressed
in Kochen and Specker's Theorem~0~\cite{kochen1}.

Conversely, there exists a plethora of propositional structures~\cite{svozil-2018-b}
that allow a partition logic interpretation, and therefore
global classical probability distributions; and yet they do not support any
faithful orthogonal representation,
and therefore no quantization and no quantum probabilities.
The simplest such example are three observables which, when depicted in a hypergraph~\cite{greechie:71,kalmbach-83,Bretto-MR3077516},
form a cyclical triangular structure.

It might not be too unreasonably to state that quantum ``contextuality'' need only to shows up if the observables
satisfy Kochen and Specker's demarcation criterion by forming some propositional structure that has no classical realization
and no joint probability distribution.
Before that one is talking about ``complementary'' configurations, which also allow global classical probability distributions
-- albeit with different probabilistic predictions yielding violations of Boole's ``conditions of possible (classical) experience''.

\section{Contextuality as object constructions}

As has been mentioned earlier, most investigations into ``contextuality'' concentrate on the second criterion (ii)
and compare discords between classical versus quantum probabilistic predictions.
Thereby a presumption is an insistence that one is only willing to accept classical Boolean propositional structures representable by (power) sets as ontological entities.

This presumption is meshed with what Bell claimed to be true: that {\it ``everything has definite properties''}~\cite{Bertlmann2020}.
That is, there is a common belief in ``Omni-definiteness'', that any outcome of some measurement reflects an ``inner property'' or ``element of physical reality''~\cite{epr}
of the ``object'' one is
pretending to ``measure''. No doubts are raised about the construction of this ``object'' which may involve important signal contributions from the measurement apparatus.
Pointedly stated: the very notion of  ``physical object''~\cite{Yanofsky-object}
-- rather than an ``image of our mind'' in the sense of Hertz~\cite{hertz-94,hertz-94e} --
may be a naive conception that is inappropriate for situations in which one is dealing with
certain types of complementary ``observables''  and, in particular,
that have no simultaneous value definiteness~\cite{pitowsky:218,2015-AnalyticKS}.
If, for instance, one would also be willing to contemplate vectors as fundamental ontological entities, then value definiteness ensues as pure states,
and arguments based on the scarcity or even absence of classical ``non-contextual'' truth assignments decay into thin air.

\begin{acknowledgments}
This research was funded in whole, or in part, by the Austrian Science Fund (FWF), Project No. I 4579-N. For the purpose of open access, the author has applied a CC BY public copyright licence to any Author Accepted Manuscript version arising from this submission.

The author declares no conflict of interest.

This paper was stimulated by a discussion on ``objectification'' and a respective draft paper~\cite{Yanofsky-object} of
Noson Yanofsky, as well as by a
question raised by Andrei Khrennikov in an email message relating to his recent paper~\cite{Khrennikov2020}.
\end{acknowledgments}

%\bibliography{svozil}


\end{document}